\documentclass[aps,pra,showpacs,twocolumn]{revtex4}
\usepackage{amssymb}
\usepackage{graphicx}
\usepackage{amsmath}
\usepackage{latexsym,epsfig}
\usepackage{float}
\usepackage{color}
\usepackage{lscape}
\usepackage{pdflscape}
\usepackage[figuresright]{rotating}

\begin{document}

\title{Estimation of BBR shift due to Stark effect for the Microwave $^{113}$Cd$^{+}$ Ion Clock}

\author{Yan-mei Yu \footnote[1]{E-mail: ymyu@aphy.iphy.ac.cn}}
\affiliation{Beijing National Laboratory for Condensed Matter Physics, Institute of Physics, Chinese Academy of Sciences, Beijing 100190,China}

\author{B. K. Sahoo \footnote[2]{E-mail: bijaya@prl.res.in}}
\affiliation{Atomic, Molecular and Optical Physics Division, Physical Research Laboratory, Navrangpura, Ahmedabad 380009, India and\\
State Key Laboratory of Magnetic Resonance and Atomic and Molecular Physics, Wuhan Institute of Physics and Mathematics,
Chinese Academy of Sciences, Wuhan 430071, China}

\date{\today}
\begin{abstract}
The microwave clock frequency of the $|5s~^2S_{1/2}, F=0,m_F=0 \rangle \leftrightarrow |5s~^2S_{1/2}, F=1,m_F=0 \rangle$ transition in the
$^{113}$Cd$^+$ ion has been reported as 15199862855.0192(10) Hz [Opt. Lett. {\bf 40}, 4249 (2015)]. Fractional systematic due to the black-body
radiation (BBR) shift ($\beta$) arising from the Stark effect in the above clock transition was used as $-1.1 \times 10^{-16}$ from our unpublished
preliminary estimation. We present here a precise value as $\beta=-1.815(77) \times 10^{-16}$ by carrying out rigorous calculations of
third-order polarizabilities of the hyperfine levels associated with the clock transition. This is determined by evaluating matrix elements of the
magnetic dipole hyperfine interaction Hamiltonian, electric dipole operator and energies between many low-lying states of $^{113}$Cd$^+$. We employ
all-order relativistic many-body theories in the frameworks of Fock-space coupled-cluster and relativistic multi-configuration Dirac-Fock methods.
\end{abstract}

\maketitle

The today's definition of unit of time in the International System of Units (SI) is based on the microwave transition frequency of the ground
state hyperfine splitting of $^{133}$Cs \cite{lombardi}. However, singly charged ions have now paved the way to carry out many precise
experiments with the advent of new technologies for cooling and trapping a singly charged ion. In fact single trapped ions $^{171}$Yb$^{+}$ \cite{huntemann}
and Al$^+$ \cite{chou} provide now clock frequencies with fractional uncertainties below $10^{-17}$. The advantage of considering an ion incur
to its nature for being controlled easily by the electromagnetic radiation. Thus, it can be isolated from the external perturbations better than
a neutral atom. As a matter of fact, several singly charged ions such as Hg$^+$, Ca$^+$, Sr$^+$, Yb$^+$ among others are under consideration in
the laboratory for atomic clocks (e.g. see \cite{bks-chapter}). However it is aimed at to construct optical frequency clocks using these ions. Both
microwave and optical frequency based clocks have advantages in their own perspectives \cite{riehle}. Since frequencies in both the categories
differ by several orders in magnitude, their applications could be totally diverse in nature. It means development of both are indispensable. Also,
it is convenient to build up microwave-frequencies using compact electronic devices than the optical frequencies that are required for the deep-space
science and commercial purposes \cite{prestage1,gill}. In fact, trapping of the Cd$^+$ ion in a semi-conductor chip has already been demonstrated
\cite{stick}. Also, cooling and crystallization of $^{113}$Cd$^+$ has been performed successfully \cite{guang} and a proof-of-principle to realize
a transportable clock using this ion has been discussed \cite{zhang}. Such technologies are useful for better quantum computers
\cite{Blinov-Nature-2004,schindler}. Micro-chip clocks are also of immense interest for various reasons. Micro-chip clocks using neutral atoms are
based on the coherent-population trapping of atoms that avoids use of microwave cavity to probe the atomic resonance \cite{gorecki}. As a
result, this provides a very compact clock. On the other hand, confinement of more number of neutral atoms can cause large collisional shifts in
the clock frequency. This can be minimized by using singly charged ions trapped in a micro-chip.

\begin{table*}[t]
\caption{Important reduced E1 matrix elements (in a.u.) between many low-lying states of Cd$^+$ from the CCSD method. $n$ corresponds to the principal
quantum number of the states. \label{tab:EDM}}
{\setlength{\tabcolsep}{12pt}
\begin{tabular}{c c c c c c c c c}\hline\hline
Transition           & $n=5$ &$n=6$ &$n=7$ &$n=8$ &$n=9$ &$n=10$ & $n=11$ &$n=12$ \\ \hline
$np~^2P_{1/2}-5s~^2S_{1/2}$  & $-1.97$  & $0.08$  & $-0.08$  & $0.07$  & $-0.05$  & 0.04   & $-0.02$   & $-0.01$   \\
$np~^2P_{3/2}-5s~^2S_{1/2}$  & $2.79$   & $-0.03$  & $0.08$  & $0.07$  & $0.06$  & $-0.05$    & $-0.03$    & $-0.02$   \\
$np~^2P_{1/2}-6s~^2S_{1/2}$  & $1.65$  & $-5.11$  & $-0.13$  &  $-0.006$ & $0.02$  & $-0.02$  & 0.02  & 0.01    \\
$np~^2P_{3/2}-6s~^2S_{1/2}$  & -2.52   & $-7.15$   & 0.35   & 0.08   & 0.02  &  $-0.005$     &   0.001  &  0.002  \\
$np~^2P_{1/2}-7s~^2S_{1/2}$  &  $-0.48$ &  3.83  &  $-9.07$    &  0.40    & $-0.08$     &  0.02     &  0.03 &  0.007   \\
$np~^2P_{3/2}-7s~^2S_{1/2}$  &  0.69     &  5.77    &  12.66    &  0.82    & 0.25     &  $-0.11$     &  $-0.05$  & $-0.02$    \\
$np~^2P_{1/2}-8s~^2S_{1/2}$  &  0.26     & $-0.96$     &  $-6.72$    & $-14.13$     &  0.74    &   $-0.19$    &  0.05  & $-0.001$    \\
$np~^2P_{3/2}-8s~^2S_{1/2}$  &  $-0.38$     & $-1.37$     & 10.08  &  $-19.67$  & $-1.42$  & 0.46  &  0.17  & 0.05    \\
$np~^2P_{1/2}-9s~^2S_{1/2}$  &  $-0.18$ & 0.51  & 1.59  & $-10.27$  & $-20.63$ & 1.17  & $-0.31$ & $-0.06$     \\
$np~^2P_{3/2}-9s~^2S_{1/2}$  &   0.25  & 0.71  &  $-2.26$ & $-15.37$ & 28.71  & $-2.17$  &  $-0.66$ & $-0.19$     \\
$np~^2P_{1/2}-10s~^2S_{1/2}$ &  0.13  & $-0.32$  & $-0.80$  & 2.36  & $-13.95$ & $-29.82$   & 1.29 & 0.26      \\
$np~^2P_{3/2}-10s~^2S_{1/2}$ &  $-0.18$  & $-0.45$ & 1.12 & 3.36 & 20.87 & 41.61  & 2.42  &  0.57     \\
$np~^2P_{1/2}-11s~^2S_{1/2}$ &  0.09 & $-0.23$  & $-0.50$  & 1.18  & $-3.38$  & 16.01  &  44.06  &  $-0.06$      \\
$np~^2P_{3/2}-11s~^2S_{1/2}$ &  $-0.13$     &  $-0.32$    &  0.70 & 1.66 & 4.85 & $-23.93$  & 61.84 & 0.37      \\
$np~^2P_{1/2}-12s~^2S_{1/2}$ &  0.09   &  $-0.22$    & $-0.44$  &  0.91  & $-2.08$   & 5.07      &  $-14.94$ & 63.89      \\
$np~^2P_{3/2}-12s~^2S_{1/2}$ &  $-0.13$   &  $0.30$    & 0.61   &  1.28  & 2.97   & $-7.39$  & $-22.16$  &  90.10    \\\hline\hline
\end{tabular}}
\end{table*}

A number of ions are under consideration for microwave clocks such as $^{199}$Hg$^+$, $^{137}$Ba$^+$, $^{171}$Yb$^+$, $^{113}$Cd$^+$, $^9$Be$^+$
etc. \cite{prestage2}. Among these ions, high precision measurement of frequency of the ground state hyperfine splitting in the $^{113}$Cd$^+$ ion
is under progress \cite{zhang2}. The advantage of this ion, except in $^{199}$Hg$^+$ \cite{berkeland}, for microwave clock is its ground state
hyperfine splitting is relatively larger; about 15.2 GHz. Moreover, it requires only one laser for cooling, pumping and detection purposes, which
is a big advantage from the experimental set-up point of view. Clock frequency of the $|F=0,M_F=0 \rangle \rightarrow |F'=1, M_F'=0\rangle $
transition in the ground state $5s~^2S_{1/2}$ of $^{113}$Cd$^+$ has already been achieved as $v_0=15199862855.0192(10)$ Hz with a fractional
precision of 6.6$\times10^{-14}$ \cite{Miao-OL-2015,Wang-OE-2013}. The major uncertainties in the systematics of these measurements come from the
Zeeman shift, but uncertainty due to the black-body radiation (BBR) shift due to stray electric fields also contribute significantly. It is
feasible to minimize uncertainties due to the Zeeman shifts in few years by improving magnetic shielding and other instrumental errors by optimizing
controlling sequences and suppressing laser technical noise \cite{Miao-OL-2015}. In case these uncertainties reach below 10$^{-16}$ level,
it would be necessary at that point to reduce uncertainty due to the BBR shift observed at the room temperature. Thus, it is imperative to estimate
this quantity more precisely for the Cd$^+$ ion. Measuring BBR shift in an atomic system at the level of 10$^{-16}$ precision level is strenuous.
Previously, a number of theoretical calculations have been performed on the BBR shifts for many of the atomic clock candidates with
ultra-high precision \cite{bks-chapter,Beloy-PRL-2006,sahoo,kallay,arora}. However, the BBR shift due to the Stark effect in the Cd$^+$ clock
transition has not been estimated so far. The fractional BBR shift used in the reported clock frequency by Miao et al \cite{Miao-OL-2015}
was based on our unpublished work. We had given them a ball-park figure of this quantity from our preliminary estimation. Therefore, knowledge of
a more reliable value of BBR shift in this clock transition is of immense interest.

The clock transition in the ground state of $^{113}$Cd$^+$ has null differential second-order dipole polarizabilities owing to the same scalar
polarizabilities of the hyperfine levels. However, the hyperfine interaction induced third-order dipole polarizabilities can offer finite
contributions to the BBR shifts in the above clock transition. In this Rapid Communication, we estimate the fractional uncertainty due to the BBR
shift of the $^{113}$Cd$^+$ clock transition by evaluating third-order dipole polarizabilities of the associated hyperfine levels. For this
purpose, we employ a relativistic coupled-cluster (RCC) method in the Fock-space framework by considering Dirac-Coulomb-Breit Hamiltonian along with
lower-order corrections due to quantum electrodynamics effects, as in our previous work \cite{yu}, and relativistic multi-configuration Dirac-Fock methods (MCDF)
method using the GRASP2K package \cite{GRASP} to evaluate the required matrix elements of the electric dipole (E1) and magnetic dipole (M1) hyperfine
interaction operators.

\begin{table}[t]
\caption{Values of the E1 matrix elements (in a.u.) for some of the transitions associated with the states having $4d^95s^2~^2D_{3/2}$ and $4d^95s5p$
configurations in Cd$^+$ from the MCDF method of GRASP2K package \cite{GRASP}. \label{tab:MCDFEEE1} }
{\setlength{\tabcolsep}{8pt}
\begin{tabular}{l c }\hline\hline											
Transition	&	 Value \\\hline		
$4d^95s^2~^2D_{3/2} - 5p~^2P_{1/2}$	&0.24 		\\	
$4d^95s^2~^2D_{3/2}-5p~^2P_{3/2}$	&0.21 			\\	
$4d^95s^2~^2D_{3/2}-6p~^2P_{1/2}$	&0.25 				\\	
$4d^95s^2~^2D_{3/2}-6p~^2P_{3/2}$ 	&0.07 				\\	
$5s~^2S_{1/2}-4d^95s(^3D)5p~^4P_{3/2}$ 	&0.28 					\\	
$5s~^2S_{1/2}-4d^95s(^3D)5p~^4P_{1/2}$ 	&0.22 					\\	
$5s~^2S_{1/2}-4d^95s(^3D)5p~^4F_{3/2}$ 	&0.03 					\\	
$5s~^2S_{1/2}-4d^95s(^3D)5p~^4D_{3/2}$ 	&0.23 					\\	
$5s~^2S_{1/2}-4d^95s(^3D)5p~^4D_{1/2}$ 	&0.80 					\\	
$5s~^2S_{1/2}-4d^95s(^3D)5p~^2P_{3/2}$ 	&-1.05 			\\	
$5s~^2S_{1/2}-4d^95s(^3D)5p~^2P_{1/2}$ 	&0.68 			\\	
$5s~^2S_{1/2}-4d^95s(^3D)5p~^2D_{3/2}$ 	&0.16 			\\	
$5s~^2S_{1/2}-4d^95s(^3D)5p~^2P_{3/2}$ 	&-1.23 					\\	
$5s~^2S_{1/2}-4d^95s(^3D)5p~^2P_{1/2}$ 	&0.67 				\\\hline\hline	
\label{E1MCDF}
\end{tabular}}
\end{table}

The light shift due to the third-order Stark effect in the hyperfine level $|F,m_F\rangle$, for $F=I+J$ with the nuclear spin $I$ and total
electron momentum $J$ and $m_F$ is the projection of $F$ on the quantization axis, of a $S$-state by a laser field with strength ${\cal E}$
can be given by
\begin{eqnarray}\label{eqs2}
\delta E_F^{(2,1)}(\omega)&=&-\frac{{\cal E}^2}{2} \alpha_F^{s(3)} .
\end{eqnarray}
Here, the third-order scalar polarizability $\alpha_F^{s(3)}$ can be conveniently expressed as \cite{Beloy-PRL-2006}
\begin{eqnarray}
\alpha^{s(3)}_F(\omega)&=&\frac{2}{3}\sqrt{(I)(I+1)(2I+1)} \Bigg\{ \begin{array}{c c c} J &I&F\\I& J&1  \end{array} \Bigg\} \\ \nonumber
                 & \times & \frac{\mu_I}{I} (-1)^{F+I+J}\left [ 2T(\omega)+C(\omega)+R(\omega) \right ],
\end{eqnarray}
where $\mu_I$ is the nuclear magnetic moment in unit of Bohr magnetron $\mu_N$. We use here $\mu_I/I=-1.2446 \mu_N$ with $I=1/2$ for
$^{113}$Cd$^+$ \cite{stones}. Other factors are given by
\begin{eqnarray}
T(\omega)&=& \frac{1}{2(2J+1)}\sum_{J',J''}\delta_{J,J''} (-1)^{J+J'} \nonumber \\
&\times& \frac{ \langle \gamma J ||{\bf D} || \gamma' J' \rangle \langle \gamma' J' ||{\bf D} || \gamma'' J'' \rangle \langle \gamma'' J'' || {\bf {\cal T}}_{hf} || \gamma J \rangle}{(E_{\gamma J}-E_{\gamma'' J''})} \nonumber \\
&\times& \Bigg[\frac{1}{(E_{\gamma J}-E_{\gamma' J'}+\omega)}+\frac{1}{E_{\gamma J}-E_{\gamma' J'}-\omega}\Bigg] ,
\label{eqtt}\\
C(\omega)&=&\frac{1}{2} \sum_{J',J''} (-1)^{J'-J''}\Bigg\{\begin{array}{c c c} 1&J& J\\ 1 & J' & J'' \end{array} \Bigg\}  \nonumber \\
&\times& \langle \gamma J ||{\bf D} || \gamma' J' \rangle \langle \gamma' J' ||{\bf  {\cal T}}_{hf} ||\gamma'' J'' \rangle \langle \gamma'' J'' ||{\bf D} || \gamma J\rangle \nonumber \\
&\times& \Bigg[\frac{1}{(E_{\gamma J}-E_{\gamma' J'} +\omega)(E_{\gamma J}-E_{\gamma'' J''}+\omega)} \nonumber \\
&+&\frac{1}{(E_{\gamma J}-E_{\gamma' J'}-\omega)(E_{\gamma J}-E_{\gamma'' J''}-\omega)}\Bigg]
\label{eqc}
\end{eqnarray}
and
\begin{eqnarray}
R(\omega)&=&\frac{1}{2(2J+1)} \langle \gamma J|| {\bf {\cal T}}_{hf} || \gamma J \rangle \sum_{J'} |\langle \gamma J || {\bf D} || \gamma' J' \rangle|^2 \nonumber \\
&\times& \left [\frac{1}{(E_{\gamma J}-E_{\gamma' J'}+\omega)^2}+\frac{1}{(E_{\gamma J}-E_{\gamma' J'}-\omega)^2}\right ],  \ \ \
\label{eqr}
\end{eqnarray}
where $E$ with subscript $\gamma J$ represent for the energies of the state with angular momentum $J$ and additional quantum numbers $\gamma$, and
${\cal T}_{hf}(r)= -i \sqrt{2} \left (\mbox{\boldmath$\alpha$} \cdot {\bf C}^{(1)}(\hat{r}) \right )/(c r^2)$ is the electronic component of
the M1 hyperfine interaction operator with the Dirac operator $\mbox{\boldmath$\alpha$}$, Racah operator $C^{(1)}$ and speed of light $c$,
for which the M1 hyperfine structure constant is defined as
\begin{equation}
A_{hf}= \frac{\mu_I}{I} \frac{\langle J\| {\bf {\cal T}}_{hf} \|J\rangle}{\sqrt{J(J+1)(2J+1)}}.
\label{eqa}
\end{equation}

\begin{table}[t]
\caption{Diagonal and off-diagonal reduced matrix elements of the ${\bf {\cal T}}_{hf}$ operator (in $10^{-8}$ a.u.) between some of the important
low-lying states of Cd$^+$ from the RCC and MCDF methods. Only the states with $4d^95s^2~^2D_{3/2}$ and $4d^95s5p$ configurations are obtained by
the MCDF method. \label{tab:HFSEOM}}
{\setlength{\tabcolsep}{5pt}
\begin{tabular}{l c   }\hline\hline
Matrix element      & Value                        \\ \hline
$5s~^2S_{1/2}-5s~^2S_{1/2}$ &228.27     \\
$5s~^2S_{1/2}-6s~^2S_{1/2}$ &$104.52$         \\
$5s~^2S_{1/2}-7s~^2S_{1/2}$ &$-65.56$            \\

$5p~^2P_{1/2}-5p~^2P_{1/2}$ &36.67           \\
$5p~^2P_{1/2}-5p~^2P_{3/2}$ &$4.14$               \\
$5p~^2P_{1/2}-6p~^2P_{1/2}$ &$19.09$            \\
$5p~^2P_{1/2}-6p~^2P_{3/2}$ &2.17                \\
$5p~^2P_{1/2}-7p~^2P_{1/2}$ &$-12.56$            \\
$5p~^2P_{1/2}-7p~^2P_{3/2}$ &$1.44$              \\

$5p~^2P_{3/2}-5p~^2P_{3/2}$ &19.66          \\
$5p~^2P_{3/2}-6p~^2P_{1/2}$ &$2.11$             \\
$5p~^2P_{3/2}-6p~^2P_{3/2}$ &$10.49$           \\
$5p~^2P_{3/2}-7p~^2P_{1/2}$ &$-1.38$            \\
$5p~^2P_{3/2}-7p~^2P_{3/2}$ &6.98             \\

$4d^95s^2 \ ^2D_{5/2} - 4d^95s^2 \ ^2D_{5/2}$             &69.73 	\\		
$4d^95s^2 \ ^2D_{3/2} - 5s~^{2}S_{1/2}$                   &$-0.03$ 	\\	
$4d^9 5s( ^3D)5p \ ^4P_{3/2}-4d^9 5s( ^3D)5p \ ^4P_{3/2}$ &135.68 	\\	
$4d^9 5s( ^3D)5p \ ^4P_{1/2}-4d^9 5s( ^3D)5p \ ^4P_{1/2}$ &63.21 	\\	
$4d^9 5s( ^3D)5p \ ^4F_{3/2}-4d^9 5s( ^3D)5p \ ^4F_{3/2}$ &$-35.59$ 	\\	
$4d^9 5s( ^3D)5p \ ^4D_{3/2}-4d^9 5s( ^3D)5p \ ^4D_{3/2}$ &115.04 	\\	
$4d^9 5s( ^3D)5p \ ^4D_{1/2}-4d^9 5s( ^3D)5p \ ^4D_{1/2}$ &$-32.53$ 	\\	
$4d^9 5s( ^3D)5p \ ^2P_{3/2}-4d^9 5s( ^3D)5p \ ^2P_{3/2}$ &212.02 	\\	
$4d^9 5s( ^3D)5p \ ^2P_{1/2}-4d^9 5s( ^3D)5p \ ^2P_{1/2}$ &11.11 	\\	
$4d^9 5s( ^3D)5p \ ^2D_{3/2}-4d^9 5s( ^3D)5p \ ^2D_{3/2}$ &87.38 	\\	
$4d^9 5s( ^1D)5p \ ^2P_{3/2}-4d^9 5s( ^1D)5p \ ^2P_{3/2}$ &73.40 	\\	
$4d^9 5s( ^1D)5p \ ^2P_{1/2}-4d^9 5s( ^1D)5p \ ^2P_{1/2}$ &44.73 	\\	
\hline\hline
\label{m1hyf}
\end{tabular}}
\end{table}

\begin{table*}[t]
\caption{Breakdown of contributions from various intermediate states $J'$ and $J''$  to $T(0)$, $C(0)$ and $R(0)$ (in $10^{-8}$ a.u.)
as defined in Eqs. (\ref{eqtt})-(\ref{eqr}). Contributions from ``Tail'', core-core, core-valence, valence-core and core contributions to
these quantities are quoted explicitly from the DHF method. Estimated uncertainties due to the matrix elements obtained
from the CCSD and MCDF methods, and contributions from the DHF method are also given. \label{tab:BBR}}
{\setlength{\tabcolsep}{5pt}
\begin{tabular}{l c |l c | l l }\hline\hline
  \multicolumn{2}{c|}{ $2T(0)$ value}      & \multicolumn{2}{c|}{ $C(0)$ value}          & \multicolumn{2}{c}{ $R(0)$ value}     \\ \hline
 & & & \\
 $J''$   & $J'=5-12p~^2P_{1/2,3/2}$   & $J''$        & $J'=5-12p~^2P_{1/2,3/2}$      & $J'$  &         \\
$6s~^2S_{1/2}$        &13350.910               &$5p~^2P_{1/2}$        &723.234                 &$5p~^2P_{1/2}$           &10300.727  \\
$7s~^2S_{1/2}$        &2319.697                &$5p~^2P_{3/2}$        &$-853.041$              &$5p~^2P_{3/2}$           &18624.610  \\
$8s~^2S_{1/2}$        &784.107                 &$6p~^2P_{1/2}$        &0.027                   &$6p~^2P_{1/2}$           &3.912     \\
$9s~^2S_{1/2}$        &374.568                 &$6p~^2P_{3/2}$        &0.006                   &$6p~^2P_{3/2}$           &0.542     \\
$10s~^2S_{1/2}$       &220.400                 &$7p~^2P_{1/2}$        &0.021                   &$7p~^2P_{1/2}$           &2.780  \\
$11s~^2S_{1/2}$       &92.335                  &$7p~^2P_{3/2}$        &$-0.016$                &$7p~^2P_{3/2}$           &2.773  \\
$12s~^2S_{1/2}$       &109.521                 &$8-12p~^2P_{1/2,3/2}$ &$-1.216$                &$8-12p~^2P_{1/2,3/2}$&7.647         \\
$[4d^95s^2~^2D_{3/2}]$ states&48.97            &$[4d^95s5p]$ states   &$-546.265$              & $[4d^95s5p]$ states     & 1349.462         \\
Uncertainty &$\pm$1087.159                     & Uncertainty          &$\pm$244.49             & Uncertainty             &$\pm$593.406   \\
 & & & \\
Tail        &142.741                           &Tail                  &$-10.640$               & Tail                &2.580          \\
Core-valence&$24.937$                          &Core-valence          & 0.417                  & Core-valence        &$-1.480$           \\
Valence-core&$-226.153$                        &Valence-core          & 0.417                  &                     &                 \\
Core-core   &21.122                            &Core-core             &$-0.188$                & Core                &429.383          \\ \hline
Final       &17263(1129)                       &Final                 &$-686(245)$             & Final               &30772(637)  \\ \hline \hline
\end{tabular}}
\end{table*}

The BBR shift due to the Stark shift in the $|F,m_F\rangle$ level can be estimated by \cite{Itano-PRA-1982}
\begin{eqnarray}\label{eqbbr}
\delta E_F^{BBR} & \simeq & -\frac{1}{2}(831.9~V/m)^2 \left (\frac{T(K)}{300} \right )^4 \alpha_F^{s(3)},
\end{eqnarray}
where $T(K)$ is the laboratory temperature in kelvin (K) with respect to the room temperature $300K$ and $\alpha_F^{s(3)}$ is in atomic
unit (a.u.). Hence, the fractional differential BBR shift  for the clock transition can be given by
\begin{eqnarray}\label{eqkbeta}
\beta_{FF'}^{BBR} &=& \frac{1}{\nu_0}(\delta E_F^{BBR} - \delta E_{F'}^{BBR}) \nonumber \\
 &\simeq& - \frac{(831.9~V/m)^2}{2\nu_0} \left (\frac{T(K)}{300} \right )^4  [\alpha_{F}^{s(3)}-\alpha_{F'}^{s(3)}] . \ \
\end{eqnarray}

Evaluation of $T$, $C$ and $R$ require many more matrix elements between different possible intermediate states of both the even and odd parities.
They mainly involve E1 matrix elements between the $ns ~ ^2S_{1/2}$ and $mp ~ ^2P_{1/2,3/2}$ states with $n$ and $m$ representing the principal
quantum numbers of different atomic states. It can also involve E1 matrix elements between the ground state and possible states with the $4d^9 ~ 5s 5p$
configurations. Moreover, it also requires E1 matrix elements involving the $4d^9 ~ 5s^2$ states with the $mp ~ ^2P_{1/2,3/2}$ states and states
with the $4d^9 ~ 5s 5p$ configurations. Similarly, both the diagonal and off-diagonal matrix elements between these states of the ${\bf {\cal T}}_{hf}$
operator will be required. Contributions to the sum of $\langle J || O_1 || J' \rangle \langle J' || O_2 || J'' \rangle \langle J'' || O_3
||J \rangle$, for $O_1$, $O_2$ and $O_3$ being either $D$ or ${\bf {\cal T}}_{hf}$ operators, from all possible excited states and core orbitals are
evaluated in several steps. The $T$ and $C$ values are evaluated by dividing into valence, core-core, core-valence, and valence-core contributions when
both $J'$ and $J''$ belong to virtual states, when both $J'$ and $J''$ belong to core orbitals, when $J'$ is from virtual and $J''$ is from core, and
when $J'$ is from core and $J''$ is from virtual, respectively. Similarly, the sum of square of the matrix element $\langle J || D || J' \rangle$ for
the evaluation of $R$ are estimated by dividing contributions as valence, core-valence and core contributions following the procedures described
in Refs. \cite{arora,jasmeet}. As can be seen from Eqs. (\ref{eqtt})-(\ref{eqr}), the low-lying excited states will contribute most to $T$, $C$
and $R$ while the core, core-valence and valence-core contributions will be relatively small. Thus, we evaluate these contributions by the
Dirac-Hartree-Fock (DHF) method. For evaluating the valence correlations, we have calculated up to $12p ~ ^2P_{1/2,3/2}$ states, that correspond to
excitation energies about 131300 cm$^{-1}$ from the ground state. All the above excited states, except those having configurations $4d^9 ~ 5s 5p$
and $4d^9 ~ 5s^2$, are calculated using the RCC method, described in our earlier works \cite{yu,bks4}, with the singles and doubles excitation
approximation (CCSD method). To determine matrix elements associated with the states with $4d^9 ~ 5s 5p$ and $4d^9 ~ 5s^2$ configurations, we
apply the MCDF method using the GRASP2K package \cite{GRASP}. Contributions above than
the $12p ~ ^2P_{1/2,3/2}$ states are estimated by the DHF method and refer to as ``Tail'' contributions.

We have compared calculated energies with the National Institute of Science and Technology database (NIST) \cite{NIST} for the states determined by the
RCC and MCDF methods, which are given in the Supplemental material. The CCSD values are found to be within 1\% accuracy and the MCDF results for
the states with $4d^9 ~ 5s 5p$ and $4d^9 ~ 5s^2$ configurations are found to be about 4\% compared to the NIST data. We also give the E1 matrix
elements among various low-lying states in Table \ref{tab:EDM} from the CCSD method. Experimental values for the E1 matrix elements of the
$5s~^2S_{1/2} - 5p~^2P_{1/2}$ and $5s~^2S_{1/2} - 5p~^2P_{3/2}$ transitions are obtained from the precise lifetime measurements of the $5p~^2P_{1/2}$ and
$5p~^2P_{3/2}$ states as 1.913(3) a.u. and 2.717(5) a.u., respectively, \cite{Moehring-PRA-2006}. Our values are in reasonably agreement with these
results. Some of the important E1 matrix elements obtained by the MCDF method are given in Table \ref{E1MCDF}. These values for the
$4d^95s^2~^2D_{3/2} - 5p~^2P_{1/2}$ and $4d^95s^2~^2D_{3/2} - 5p~^2P_{3/2}$ transitions are inferred from experiments as 0.487(17) a.u. and 0.295(37)
a.u., respectively, \cite{Xu-PRA-2004}. They are almost twice than the calculated values. The remaining E1 matrix elements from the MCDF method can
be found from the Supplemental material.

Similarly, the reduced matrix elements of ${\bf {\cal T}}_{hf}$ obtained in the RCC and MCDF methods for some of the important contributing states
are listed in Table \ref{m1hyf}. The rest of the matrix elements for other excited states that are calculated by us are given in the Supplemental
material. Using the relation given by Eq. (\ref{eqa}), we have also extracted out the reduced expectation values for the $5s~^2S_{1/2}$, $5p~^2_{3/2}$ and
$4d^95s^2 \ ^2D_{5/2}$ states as $227.66(1)\times 10^{-8}$ a.u. \cite{Miao-OL-2015}, $18.94\times 10^{-8}$ a.u. \cite{Tanaka-PRA-1996} and
$75.10(3) \times 10^{-8}$ a.u. \cite{hermann}, respectively, from the precise measurements of $A_{hf}$. Comparison of these values with our
calculations show good agreement between them.

Using all the reduced matrix elements discussed above, we determine the static values of $T$, $C$ and $R$ for the ground state of $^{113}$Cd$^+$
and give them in Table \ref{tab:BBR} after summing contributions from various intermediate states. To reduce uncertainties in these calculations,
we use the experimental energies and replace the calculated reduced matrix elements by the precisely known experimental values wherever possible.
It to be noted that knowledge of signs of the matrix elements are necessary for the evaluation of $T(0)$ and $C(0)$. However, signs of the experimental
E1 amplitudes cannot be determined. In this case, we use the magnitudes from the measurements and signs from the MCDF results. We also give contributions
explicitly from the core-core, core-valence, valence-core and core contributions to these quantities in the same table. As can be seen from the table,
the $6s~^2S_{1/2}$ and $7s~^2S_{1/2}$ states contribute extremely large to $T(0)$. Other $S$-states also contribute significantly here. Tail, core-valence
and valence-core contributions to $T(0)$ are also non-negligible, but core-core contribution is found to be small. The net contributions from the states
with $4d^95s^2~^2D_{3/2}$ configurations are also found to be relatively small. The dominant uncertainties come from the matrix elements involving the
calculated $S$-states. On other hand, most of the contributions to $C(0)$ comes from the $5P$-states. The net contributions from the states with
$4d^95s5p$ configurations are also found to be quite significant. For the $R(0)$ value, the most dominant contributions come from the $5P$-states followed
by the states with $4d^9 ~ 5s 5p$ configuration. The core contribution is also found to be quite large. We estimate uncertainties to the above quantities
by assigning an overall $\pm$3\% errors to the CCSD results and $\pm$20\% errors to the MCDF results. Uncertainties due to the other contributions given
from the DHF method are estimated to be about 10\%.

Accounting for all these values along with the uncertainties, we find the differential BBR shift coefficient $\beta_{FF'}^{BBR}=-1.815(77)\times10^{-16}$
for the $|5s~^2S_{1/2}, F=0,m_F=0 \rangle \leftrightarrow |5s~^2S_{1/2},F'=1,m_{F'}=0 \rangle$ clock transition in the $^{113}$Cd$^+$ ion in contrast to
$-1.1\times10^{-16}$ that was used in Ref. \cite{Miao-OL-2015}. Uncertainty to this value can be further reduced by improving precision of the matrix elements
obtained by the MCDF method.

We thank Dr. J. W. Zhang, Joint Institute for Measurement Science, Tsinghua University, China for many useful discussions and providing us experimental
information. Y.Y. is supported by the National Natural Science Foundation of China under Grant No. 91536106, the CAS XDB21030300, and the NKRD Program of China
(2016YFA0302104). B.K.S. acknowledges financial support from CAS through the PIFI fellowship under the project number 2017VMB0023 and partly by the TDP project
of Physical Research Laboratory (PRL), Ahmedabad and the computations were carried out using the Vikram-100 HPC cluster of PRL.

\end{document}


\section*{Supplemental Material}

Many excited states of Cd$^+$ ion along with its ground state can be conveniently calculated by treating configurations of these states as a common core $[4d^{10}]$ with different valence orbitals.
In the relativistic coupled-cluster theory (RCC) {\it ansatz}, we express wave functions of states with a closed-core and a valence orbital $v$  as \cite{bks1,bks2}
\begin{eqnarray}
 |\Psi_v \rangle = e^T \{1+S_v \} |\Phi_v \rangle,
 \label{eqcc}
\end{eqnarray}
where $|\Phi_v \rangle$ is the reference state and $T$ and $S_v$ are the excitation operators involving core and core-valence electrons, respectively, that produce different
determinantal states from $|\Phi_v \rangle$. Again, we define $|\Phi_v \rangle = a_v^{\dagger}|\Phi_0\rangle$, where $|\Phi_0\rangle$ is a mean-field wave function of the
closed-core $[4d^{10}]$. We apply the Dirac-Hartree-Fock (DHF) method by considering the Dirac-Coulomb-Breit Hamiltonian (DCB) along with lower-order corrections due to quantum
electrodynamics (QED) effects to obtain the mean-field wave function as described in Refs. \cite{bks4,yu}. We consider all the core electrons active and approximate the RCC method
only to the singles and doubles excitations (CCSD method). Amplitudes of the RCC excitation operators are evaluated using the equations
\begin{eqnarray}
 \langle \Phi_0^* \vert \overline{H}_N  \vert \Phi_0 \rangle &=& 0
\label{eqt}
 \end{eqnarray}
and
\begin{eqnarray}
 \langle \Phi_v^* \vert \big ( \overline{H}_N - \Delta E_v \big ) S_v \vert \Phi_v \rangle &=&  - \langle \Phi_v^* \vert \overline{H}_N \vert \Phi_v \rangle ,
\label{eqsv}
\end{eqnarray}
where $\vert \Phi_0^* \rangle$ and $\vert \Phi_v^* \rangle$ are the excited state configurations with respect to the DHF states $\vert \Phi_0 \rangle$ and $\vert \Phi_v \rangle$ respectively
and $\overline{H}_N= \big ( H_N e^T \big )_l$ with subscript $l$ represents for the linked terms only. The attachment energy $\Delta E_v$ of the electron in the valence orbital $v$ is
evaluated by
\begin{eqnarray}
 \Delta E_v  = \langle \Phi_v \vert \overline{H}_N \left \{ 1+S_v \right \} \vert \Phi_v \rangle .
 \label{eqeng}
\end{eqnarray}

After obtaining amplitudes of the RCC operators using the above described equations, the E1 matrix elements and the matrix elements of the ${\bf {\cal T}}_{hf}$ operator between the states
$\vert \Psi_i \rangle$ and $\vert \Psi_f \rangle$ are evaluated using the expression
\begin{eqnarray}
\langle \Psi_f| O | \Psi_i\rangle &=& \frac{\langle\Phi_f|\tilde{O}_{fi}|\Phi_i\rangle}{\sqrt{\langle\Phi_f|\{1+\tilde{N}_f\}|\Phi_f\rangle
\langle\Phi_i|\{1+\tilde{N}_i\}|\Phi_i \rangle}} , \nonumber \\
\label{eqno}
\end{eqnarray}
where $\tilde{O}_{fi}=\{1+S_f^{\dagger} \} e^{T^{\dagger}} O e^T \{1+S_{i}\}$ and $\tilde{N}_{k=f,i}=\{1+S_k^{\dagger} \} e^{T^{\dagger}}e^T \{1+S_{k}\}$, assuming either $O\equiv D$ or
$O \equiv {\bf {\cal T}}_{hf}$. The expectation values are evaluated by assuming $\vert \Psi_i \rangle = \vert \Psi_f \rangle = \vert \Psi_v \rangle$. As can be seen, the above
expression involves two non-terminating series in the numerator and denominator, which are $e^{T^{\dagger}} O e^T$ and $e^{T^{\dagger}} e^T$ respectively. We adopt iterative procedures
to account for contributions from these non-terminating series as have been described in our earlier works \cite{bks2,bks4,bks3}.

The above RCC method formulated in the Fock space formalism is not suitable to obtain the low-lying states with the $4d^9 ~ 5s 5p$ and $4d^9 ~ 5s^2$ configurations of Cd$^+$ from a
common reference state along with the other excited states. Since matrix elements involving these states are less significant to the properties of our interest, we calculate atomic
state functions (ASFs) of these states by employing relativistic multi-configuration Dirac-Hartree-Fock (MCDF) method that is
available in the GRASP2K package \cite{GRASP}. In this MCDF method, we again consider only the singles and doubles excitations to construct configuration state functions (CSFs). We have
also taken into account the Breit and QED interactions that are available in this package. Due to availability of limited computational resources with us, we only include upper core
orbitals in the active space of the MCDF method that are summarized in Table \ref{tab:csf}.

In Table \ref{tab:energy}, we present excitation energies of the excited states of Cd$^+$ from both the CCSD and MCDF methods. These values are also compared with the data listed in the National
Institute of Science and Technology database \cite{NIST}. Small differences between the CCSD results and NIST data suggest that our calculations are within 1\% accuracy. Energies obtained
by the MCDF method have relatively large errors. Nonetheless, we use the experimental energies to reduce uncertainties in the evaluation of the polarizabilities.

Similarly, a list of important diagonal and off-diagonal matrix elements of ${\bf {\cal T}}_{hf}$ obtained by the CCSD and MCDF methods are given in atomic unit (a.u.) in Table \ref{tab:HFSEOM}.
Some of the diagonal elements extracted from the measured magnetic dipole hyperfine structure constants are also quoted in the same table. It shows very good agreement between the measured and
calculated values; especially from the CCSD method. In our polarizability calculations, we replace calculated values with the available experimental results to reduce the uncertainties. Since the
higher excited states than the $4d^95s^2 \ ^2D_{5/2}$ state that are obtained by the MCDF method do not contribute significantly to the polarizabilities, accuracies of the polarizabilities
obtained by using these matrix elements are estimated to be within the intended level.

The E1 matrix elements obtained from the CCSD and MCDF methods are quoted in Table \ref{tab:MCDFEEE1}. Experimental values of the E1 matrix elements of few low-lying transitions are also quoted by
extracting them out from the precise lifetime measurements of the atomic states in Cd$^+$. As can be seen values obtained by the CCSD method are more reliable. The E1 matrix elements of the
$4d^95s^2 \ ^2D_{3/2}-4d^{10}5s \ ^2P_{1/2}$ and $4d^95s^2 \ ^2D_{3/2}-4d^{10}5s \ ^2P_{5/2}$ transitions obtained by the MCDF method are found to be almost half of the experimental values.
In this case, we use the experimental values in our polarizability calculations but signs of these values are decided from the MCDF results. Again, contributions from the remaining E1 matrix
elements of the MCDF method are accounted for assuming 20\% uncertainties in these values.

\begin{table*}[]
\caption{A brief summary of active CSFs considered in our calculations using the MCDF method. As seen orbitals from the inner core $[3d^{10}]$ are not included in the MCDF procedure due to limitations in the
computational resources. Different possible multi-reference (MR) space for constructing atomic state functions (ASFs) and higher orbitals up to which active space is constructed for the MCDF
method are mentioned. Total number of CSFs generated in these process are quoted as $N_{CSFs}$.\label{tab:csf} }
{\setlength{\tabcolsep}{8pt}
\begin{tabular}{l | l | l | l }\hline\hline
ASFs                                        &   MR space                                         & Active space      & $N_{CSFs}$ \\ \hline
$4d^{10}5s$,$4d^95s^2$,$4d^{10}6s$  & \{$4s^{2}$\}S\{$4p^{6}$\}\{$4d^{10}5s$\}SD    & \{8s,6p,6d,5f\} &59636        \\
                                    & \{$4s^{2}$\}S\{$4p^{6}$\}\{$4d^{10}6s$\}SD    &                 &            \\
                                    & \{$4s^{2}$\}S\{$4p^{6}$\}\{$4d^{9}5s^2$\}SD   &                 &            \\
                                    & \{$4s^{2}$\}\{$4p^{6}$\}S\{$4d^{10}5s$\}SD    &                 &            \\
                                    & \{$4s^{2}$\}\{$4p^{6}$\}S\{$4d^{10}6s$\}SD    &                 &            \\
                                    & \{$4s^{2}$\}\{$4p^{6}$\}S\{$4d^{9}5s^2$\}SD   &                 &            \\ \hline
$4d^{10}5p$,$4d^95s5p$,$4d^{10}6p$  &\{$4s^{2}$\}S\{$4p^{6}$\}\{$4d^{10}5p$\}SD     & \{7s,8p,6d,5f\} &480071       \\
                                    &\{$4s^{2}$\}S\{$4p^{6}$\}\{$4d^{10}6p$\}SD     &                 &         \\
                                    &\{$4s^{2}$\}S\{$4p^{6}$\}\{$4d^{9}5s5p$\}SD    &                 &         \\
                                    &\{$4s^{2}$\}\{$4p^{6}$\}S\{$4d^{10}5p$\}SD     &                 &         \\
                                    &\{$4s^{2}$\}\{$4p^{6}$\}S\{$4d^{10}6p$\}SD     &                 &         \\
                                    &\{$4s^{2}$\}\{$4p^{6}$\}S\{$4d^{9}5s5p$\}SD    &                 &         \\\hline\hline
\end{tabular}}
\end{table*}

\begin{table*}[h]
\caption{Excitation energies (in cm$^{-1}$) of many low-lying states of Cd$^{+}$ from the CCSD and MCDF methods. Calculated values are compared against the data listed in the NIST database
\cite{NIST}. Comparison among the calculations with the NIST data suggests CCSD values are less than 1\% accurate while MCDF results have relatively large uncertainties. \label{tab:energy}}
{\setlength{\tabcolsep}{6pt}
\begin{tabular}{ l c c c  }\hline\hline
Level	   &        CCSD	&  MCDF     & NIST  \\ \hline
$5p~^2P_{1/2}$ &  44145.46	& 43005    &44136.08    \\
$5p~^2P_{3/2}$ &  46648.51	& 45165     &  46618.55  \\
$6s~^2S_{1/2}$ &  82721.43	& 79291    &82990.66  \\
$6p~^2P_{1/2}$ &  94488.98	& 90686  &  94710.41   \\
$6p~^2P_{3/2}$ &  95181.70	& 91331  &  95383.63  \\
$7s~^2S_{1/2}$ &  107002.20	& &107300.88  \\
$7p~^2P_{1/2}$ &  111960.45	& &112361.06  \\
$7p~^2P_{3/2}$ &  112261.59	& &112490.39  \\
$8s~^2S_{1/2}$ &  117760.13	& &118040.65  \\
$8p~^2P_{1/2}$ &  120302.71	& &120618.46  \\
$8p~^2P_{3/2}$ &  120461.36	& &120711.18  \\
$9s~^2S_{1/2}$ &  123519.91	& &123751.48  \\
$9p~^2P_{1/2}$ &  124969.52	& &125223.06  \\
$9p~^2P_{3/2}$ &  125062.43	& &125254.49  \\
$10s~^2S_{1/2}$ & 127031.96	& &127152.89  \\
$10p~^2P_{1/2}$ & 127862.19	& &127996.73  \\
$10p~^2P_{3/2}$ & 127916.41	& &128076.30  \\
$11s~^2S_{1/2}$ & 129423.70	& &129343.05  \\
$11p~^2P_{1/2}$ & 129830.45	& &129956.30  \\
$11p~^2P_{3/2}$ & 129855.73	& &130216.60 \\
$12s~^2S_{1/2}$ & 131192.37	& &130835.86 \\
$12p~^2P_{1/2}$ & 131333.96	& &131261.10 \\
$12p~^2P_{3/2}$ & 131341.95	& &131374.70 \\	
$4d^95s^2~^2D_{5/2}$	    &	& 69035 	&	69258.51	 \\	
$4d^95s^2~^2D_{3/2}$	    &	& 74680 	&	74893.66		\\	
$4d^95s(^3D)5p~^4P_{3/2}$	&	& 104981 	&	109440.86		 \\	
$4d^95s(^3D)5p~^4P_{1/2}$	&	& 107886 	&	112196.5	 	 \\	
$4d^95s(^3D)5p~^4F_{3/2}$	&	& 108486 	&	112785.19		 \\	
$4d^95s(^3D)5p~^4D_{3/2}$	&	& 116263	&	116225.71	 	 \\	
$4d^95s(^3D)5p~^4D_{1/2}$	&	& 115311  &	117988.79		 \\	
$4d^95s(^3D)5p~^2P_{3/2}$	&	& 113909	&	119056.11 	 \\	
$4d^95s(^3D)5p~^2P_{1/2}$	&	& 115086 	&	119293.99	 	 \\	
$4d^95s(^3D)5p~^2D_{3/2}$	&	& 112390 	&	120135.22 	\\	
$4d^95s(^1D)5p~^2P_{3/2}$	&	& 129150 	&	129344.86 	 \\
\hline\hline
\end{tabular}}
\end{table*}

\begin{table*}[]
\caption{Diagonal and off-diagonal reduced matrix elements of the magnetic hyperfine operator ${\bf {\cal T}}_{hf}$ (in $10^{-8}$ a.u.) from the CCSD and MCDF methods. Some of the diagonal elements
are also compared with the available experimental results. \label{tab:HFSEOM}}
{\setlength{\tabcolsep}{5pt}
\begin{tabular}{l c c c   }\hline\hline
Matrix element      & CCSD & MCDF  & Experiment                           \\ \hline
$5s~^2S_{1/2}-5s~^2S_{1/2}$ &228.27   & 221.12 & 227.66(1) \cite{Miao-OL-2015}  \\
$6s~^2S_{1/2}-6s~^2S_{1/2}$ & 47.82   & 39.13  &                           \\
$5s~^2S_{1/2}-6s~^2S_{1/2}$ &$104.52$ & 92.99  &        \\
$5s~^2S_{1/2}-7s~^2S_{1/2}$ &$-65.56$ &           \\
$5s~^2S_{1/2}-8s~^2S_{1/2}$ &44.98    &           \\
$5s~^2S_{1/2}-9s~^2S_{1/2}$ &$-33.88$ &           \\
$5s~^2S_{1/2}-10s~^2S_{1/2}$&25.94    &           \\
$5s~^2S_{1/2}-11s~^2S_{1/2}$&20.44   &           \\
$5s~^2S_{1/2}-12s~^2S_{1/2}$&20.95    &           \\

$5p~^2P_{3/2}-5p~^2P_{3/2}$ &19.66    &  18.58 & 18.94 \cite{Tanaka-PRA-1996}        \\
$5p~^2P_{3/2}-6p~^2P_{1/2}$ &$2.11$    &   1.78  &     \\
$5p~^2P_{3/2}-6p~^2P_{3/2}$ &$10.49$   &  9.17  &      \\

$6p~^2P_{1/2}-6p~^2P_{1/2}$ &9.95     &   7.73 &       \\
$6p~^2P_{1/2}-6p~^2P_{3/2}$ &1.05     &   0.60 &      \\

$6p~^2P_{3/2}-6p~^2P_{3/2}$ &5.62     &    4.65 &     \\

$4d^95s^2 \ ^2D_{3/2} - 4d^95s^2 \ ^2D_{3/2}$             	&	& 99.21 	&  75.10(3) \cite{hermann}		\\	
$4d^95s^2 \ ^2D_{3/2} - 5s~^{2}S_{1/2}$                  	&	& $-0.03$ 	 		 &		\\	
$4d^9 5s( ^3D)5p \ ^4P_{3/2} - 4d^9 5s( ^3D)5p \ ^4P_{3/2}$	&	& 135.68 		 		 &		\\	
$4d^9 5s( ^3D)5p \ ^4P_{1/2} - 4d^9 5s( ^3D)5p \ ^4P_{1/2}$ &	& 63.21 				 &		\\	
$4d^9 5s( ^3D)5p \ ^4F_{3/2} - 4d^9 5s( ^3D)5p \ ^4F_{3/2}$	&	& $-35.56$ 				 &		\\	
$4d^9 5s( ^3D)5p \ ^4D_{3/2} - 4d^9 5s( ^3D)5p \ ^4D_{3/2}$	&	& 115.04 		 		 &		\\	
$4d^9 5s( ^3D)5p \ ^4D_{1/2} - 4d^9 5s( ^3D)5p \ ^4D_{1/2}$ &	& $-32.53$ 	 			 &		\\	
$4d^9 5s( ^3D)5p \ ^2P_{3/2} - 4d^9 5s( ^3D)5p \ ^2P_{3/2}$	&	& 73.40 	   		 &		\\	
$4d^9 5s( ^3D)5p \ ^2P_{1/2} - 4d^9 5s( ^3D)5p \ ^2P_{1/2}$ &	& 44.73 				 &		\\	
$4d^9 5s( ^3D)5p \ ^2D_{3/2} - 4d^9 5s( ^3D)5p \ ^2D_{3/2}$ &	& 87.38 		 		 &		\\	
$4d^9 5s( ^1D)5p \ ^2P_{3/2} - 4d^9 5s( ^1D)5p \ ^2P_{3/2}$ &	& 212.02 		 		 &		\\	
$4d^9 5s( ^1D)5p \ ^2P_{1/2} - 4d^9 5s( ^1D)5p \ ^2P_{1/2}$ &	& 11.11 	 			 &		\\
$4d^9 5s( ^3D)5p \ ^4P_{3/2} - 4d^{10} 5p~^{2}P_{1/2}$      &	& $-6.97$ 	 			 &	    \\	
$4d^9 5s( ^3D)5p \ ^4P_{3/2} - 4d^{10} 5p~^{2}P_{3/2}$     	&	& $-11.59$ 		 		 &		\\	
$4d^9 5s( ^3D)5p \ ^4P_{1/2} - 4d^{10} 5p~^{2}P_{1/2}$     	&	& 11.14 	 			 &		\\	
$4d^9 5s( ^3D)5p \ ^4P_{1/2} - 4d^{10} 5p~^{2}P_{3/2}$      &	& 10.29 				 &		\\	
$4d^9 5s( ^3D)5p \ ^4F_{3/2} - 4d^{10} 5p~^{2}P_{1/2}$      &	& $-0.94$ 	    	 	 &	  	\\	
$4d^9 5s( ^3D)5p \ ^4F_{3/2} - 4d^{10} 5p~^{2}P_{3/2}$      &	& $-2.01$ 		 		 &		\\	
$4d^9 5s( ^3D)5p \ ^4D_{3/2} - 4d^{10} 5p~^{2}P_{1/2}$      &	& 2.74 	 		 &		\\	
$4d^9 5s( ^3D)5p \ ^4D_{3/2} - 4d^{10} 5p~^{2}P_{3/2}$     	&	& $-6.03$ 				 &		\\	
$4d^9 5s( ^3D)5p \ ^2D_{3/2} - 4d^{10} 5p~^{2}P_{1/2}$     	&	& 3.11 		 		 &		\\
$4d^9 5s( ^3D)5p \ ^4D_{1/2} - 4d^{10} 5p~^{2}P_{1/2}$      &	& $-8.04$ 	 			 &		\\	
$4d^9 5s( ^3D)5p \ ^4D_{1/2} - 4d^{10} 5p~^{2}P_{3/2}$     	&	& $-10.02$ 	        	 &		\\	
$4d^9 5s( ^3D)5p \ ^2P_{3/2} - 4d^{10} 5p~^{2}P_{1/2}$      &	& $-1.51$ 				 &		\\
$4d^9 5s( ^3D)5p \ ^2P_{3/2} - 4d^{10} 5p~^{2}P_{3/2}$     	&	& 3.29 	 			 &		\\	
$4d^9 5s( ^3D)5p \ ^2P_{1/2} - 4d^{10} 5p~^{2}P_{1/2}$     	&	& $-1.71$ 	 			 &		\\	
$4d^9 5s( ^3D)5p \ ^2P_{1/2} - 4d^{10} 5p~^{2}P_{3/2}$      &	& $-1.73$ 	 			 &		\\	
$4d^9 5s( ^3D)5p \ ^2D_{3/2} - 4d^{10} 5p~^{2}P_{3/2}$      &	& 4.39 	 			 &		\\	
$4d^9 5s( ^1D)5p \ ^2P_{3/2} - 4d^{10} 5p~^{2}P_{1/2}$      &	& $-12.64$ 	 			 &		\\	
$4d^9 5s( ^1D)5p \ ^2P_{3/2} - 4d^{10} 5p~^{2}P_{3/2}$     	&	& 9.99 	 			 &		\\	
$4d^9 5s( ^1D)5p \ ^2P_{1/2} - 4d^{10} 5p~^{2}P_{1/2}$     	&	& $-5.83$ 	 			 &		\\	
$4d^9 5s( ^1D)5p \ ^2P_{1/2} - 4d^{10} 5p~^{2}P_{3/2}$     	&	& $-4.03$ 	 	 		 &		\\ \hline\hline	
\end{tabular}}
\end{table*}

\begin{table*}[]
\caption{Reduced E1 matrix elements (in a.u.) obtained by the CCSD and MCDF methods. We have also inferred these values from the precisely available lifetime measurements of some of the
low-lying excited state to compare with the calculated values. The CCSD results are found to be more accurate than the MCDF values, but the calculated values serve to decide the signs of the matrix
elements to use the experimental values in the evaluation of polarizabilities. \label{tab:MCDFEEE1} }
{\setlength{\tabcolsep}{8pt}
\begin{tabular}{l c  c  c  }\hline\hline												
Transition	                                        &	CCSD & MCDF      &	Experiment		 \\\hline	
$5s~^2S_{1/2}-5p~^2P_{1/2}$	                        &	$-1.97$ & $-1.98$ &	1.913(3) \cite{Moehring-PRA-2006}	 	\\
$5s~^2S_{1/2}-5p~^2P_{3/2}$      	                &	2.79 &  2.82	&	2.717(5) \cite{Moehring-PRA-2006}	 	 \\	
$6s~^2S_{1/2}-5p~^2P_{3/2}$	                        &	2.52 &  2.58 	&	            	  	\\
$6s~^2S_{1/2}-5p~^2P_{1/2}$	                        &	1.65 & 1.70 	&	           	  	\\
$6p~^2P_{1/2}-6s~^2S_{1/2}$	                        &	$-5.11$ & $-5.40$ 		            	\\
$6p~^2P_{3/2}-6s~^2S_{1/2}$	                        &	$-7.15$ & $-7.57$ 	            	  	\\
$5s~^2S_{1/2}-6p~^2P_{1/2}$	                        &	0.08 & 0.07 	  &	            	  	\\
$6p~^2P_{3/2}-5s~^2S_{1/2}$	                        &	0.03 & 0.02 	  &	            	  	\\
$4d^95s^2~^2D_{3/2}-5p~^2P_{1/2}$	                &	& 0.24	& 	   	0.487(17)  \cite{Xu-PRA-2004}	  \\	
$4d^95s^2~^2D_{3/2}-5p~^2P_{3/2}$	                &	& 0.11 	& 	   	0.295(37)  \cite{Xu-PRA-2004}	\\	
$4d^95s^2~^2D_{3/2}-6p~^2P_{1/2}$	                &	& 0.25	& 	   		 		\\	
$4d^95s^2~^2D_{3/2}-6p~^2P_{3/2}$ 	                &	& 0.07 	& 	   		 		\\	
$5s_{1/2}~^2S_{1/2}-4d^95s(^3D)5p~^4P_{3/2}$ 	 	&	& 0.28 	&      				\\	
$5s_{1/2}~^2S_{1/2}-4d^95s(^3D)5p~^4P_{1/2}$ 	 	&	& 0.22 	&      				\\	
$5s_{1/2}~^2S_{1/2}-4d^95s(^3D)5p~^4F_{3/2}$ 	 	&	& 0.03 	&      				\\	
$5s_{1/2}~^2S_{1/2}-4d^95s(^3D)5p~^4D_{3/2}$ 	 	&	& 0.23    &	   				\\	
$5s_{1/2}~^2S_{1/2}-4d^95s(^3D)5p~^4D_{1/2}$ 	 	&	& 0.80 	&      				\\	
$5s_{1/2}~^2S_{1/2}-4d^95s(^3D)5p~^2P_{3/2}$ 	 	&	& $-1.05$        &				\\	
$5s_{1/2}~^2S_{1/2}-4d^95s(^3D)5p~^2P_{1/2}$ 	 	&	& 0.68	     &	 	     	\\	
$5s_{1/2}~^2S_{1/2}-4d^95s(^3D)5p~^2D_{3/2}$ 	 	&	& 0.16          &	 	       \\	
$5s_{1/2}~^2S_{1/2}-4d^95s(^1D)5p~^2P_{3/2}$ 	    &	& $-1.23$ 	      &				\\	
$5s_{1/2}~^2S_{1/2}-4d^95s(^1D)5p~^2P_{1/2}$ 	    &	& 0.67 	      &				\\\hline\hline	
\end{tabular}}
\end{table*}